%% file: paper.tex
\documentclass[conference]{IEEEtran}
\IEEEoverridecommandlockouts
\usepackage{multirow}
\usepackage{verbatim} 
\ifCLASSINFOpdf
   \usepackage[pdftex]{graphicx}
   \graphicspath{{images/}}
   \DeclareGraphicsExtensions{.eps}
\else
\fi
\usepackage[cmex10]{amsmath}
\usepackage{array}

\usepackage{caption}
\usepackage[caption=false,font=footnotesize]{subfig}
\usepackage{fixltx2e}

\usepackage{float}
\usepackage{stfloats}

\usepackage{url}

\begin{document}

%
\title{Asynchronous Rumour Spreading in Social and Signed Topologies}


\author{
  \IEEEauthorblockN
  {
    Christos Patsonakis,
    Mema Roussopoulos
  }
  \IEEEauthorblockA
  {
    Department of Informatics and Telecommunications\\
    National and Kapodistrian University of Athens \\
    Athens, Greece\\
    \{c.patswnakis,mema\}@di.uoa.gr
  }
}


%


\maketitle

\begin{abstract}

In this paper, we present an experimental analysis of the asynchronous \emph{push \& pull} 
rumour spreading protocol.  This protocol is, to date, the best-performing rumour spreading
protocol for simple, scalable, and robust information dissemination in distributed systems.
We analyse the effect that multiple parameters have on the protocol's performance, such as
using memory to avoid contacting the same neighbor twice in a row, varying the stopping
criteria used by nodes to decide when to stop spreading the rumour, employing more sophisticated
neighbor selection policies instead of the standard uniform random choice, and others.

Prior work has focused on either providing theoretical upper bounds regarding the number of 
rounds needed to spread the rumour to all nodes, or, proposes improvements by adjusting isolated parameters. 
To our knowledge, our work is the first to study how \emph{multiple} parameters affect system behaviour
both in \emph{isolation and combination} and
under a wide range of values. 

Our analysis is based on experimental simulations using real-world
social network datasets, thus complementing prior theoretical work to shed light on how the protocol behaves
in practical, real-world systems.  We also study the behaviour of the protocol on a special type of
social graph, called signed networks (e.g. Slashdot and Epinions), whose links indicate
stronger trust relationships.
Finally, through our detailed analysis, we demonstrate how a few simple additions to the protocol can improve 
the total time required to inform 100\% of the nodes by a maximum of 99.69\% and an average of 86.04\%.

\end{abstract}


\IEEEpeerreviewmaketitle

\input{introduction}
\input{background}
\input{relatedwork}
\input{methodology}
\input{experimentsandresults}
\input{conclusionandfuture}

\bibliographystyle{IEEEtran}
\bibliography{IEEEabrv,bibliography}

%


\end{document}

%% file: introduction.tex
\section{Introduction}
\label{sect:intro}
Efficient information dissemination is a fundamental problem in distributed systems.
In particular, rumour spreading is a class of randomized dissemination protocols that
have been proposed for a variety of distributed applications such as maintaining
consistency in replicated database settings~\cite{alan.demers}, multicast~\cite{birman1999},
distributed ranking~\cite{chiuso2011}, and others.

Rumour spreading protocols are well-known
for their robustness, simplicity, and scalability properties \cite{pat.eugster}.
In distributed systems that use rumour spreading,
information may be spread in one of three ways:  1) by having the \emph{informed} nodes (those with the information
to be disseminated) actively \emph{push}
the information to the rest of the network, 2) by having the \emph{uninformed} nodes request, or \emph{pull},
the information, or 3) by combining both \emph{push \& pull} approaches. Nodes can engage 
in the protocol in complete synchrony by pushing and/or pulling at
the same time (in rounds), 
or asynchronously, with each node
running its own clock.  The goal is to have the \emph{rumour} (the information of interest) propagate quickly and
efficiently throughout the system.

While there is a wealth of prior work on rumour spreading protocols, the existing literature 
has gaps in two important areas.
First, the spectrum of possible improvements to the plain vanilla push/pull versions of these protocols
is large and to date, remains mostly unexplored. 
There are several factors that affect the performance of these protocols, typically gauged as how quickly
information is propagated throughout the system. Prior work is largely theoretical in nature and/or proposes
improving performance by adjusting isolated parameters of the protocol such as, the choice of nodes to neighbor
with to achieve a more efficient topology of inter-connections between nodes, the choice of neighbor to which to
send the rumour, the use of memory to avoid sending the rumour twice to the same neighbor
and stopping criteria to avoid propagating the rumour more than needed.
To date, {\bf no} study has focused on studying how \emph{multiple} parameters affect the protocol's behaviour 
both in {\emph{isolation and combination} and under a wide range of values. 

Second, while prior work examines the behaviour of rumour spreading protocols in a variety of topologies (e.g.,
meshes and tori~\cite{Juurlink1996}, butterfly networks~\cite{Sibeyn2005}), very few publications
focus on social topologies. 
There is a wealth of diverse distributed applications that link nodes based on
trust relationships, or, have interconnection graphs that exhibit power-law,
scale-free and small-world properties, thus forming social, or, social-like
topologies \cite{alan.mislove0}. Social-based file sharing \cite{tribler}, viral marketing \cite{marketing}
and ensuring eventual consistency amongst the sites of a replicated database (\cite{alan.demers}, \cite{complexsocial}),
are only a few examples of applications that use rumour spreading in these
topologies and can benefit from our findings.

Until now, nearly all of the published results regarding rumour spreading protocols on social graphs
are theoretical analysis.  While they provide upper bounds on the number of rounds
that these protocols require to achieve propagation, they do not provide any insight with regards to
network-related metrics.  Such information includes measured time delays and
the network load incurred by the exchange of protocol messages. Moreover, a substantial portion of the
published results focuses only on synchronous rumour spreading.

In this paper, we provide an in-depth experimental analysis of the \emph{asynchronous push \& pull} 
rumour spreading protocol. We focus on this particular type of protocol  
since it is known to combine the benefits of both \emph{push} and \emph{pull} (\cite{alan.demers, richard.karp}).
Furthermore, it has recently been proven that the asynchronous variant 
performs substantially better than the synchronous in both preferential attachment graphs
(PA - synthetic graphs that resemble social graphs) and real-world social networks (\cite{bj.doer0},~\cite{bj.doer2}).

Our methodology for extracting results is via simulations using a variety of real-world social network
datasets (ten in total).  
We evaluate the impact of individual parameters on protocol performance over a wide 
range of values.  We also present combined experiments, where we leverage the knowledge acquired
from our single-parameter experiments to illustrate how an \emph{Enhanced push \& pull} protocol with an
intelligent parameter selection can significantly outperform the plain vanilla push and pull protocol.
Moreover, we explore how information flows on a special type of social
networks, called signed networks, whose links
indicate stronger trust values than those of common
social networks. We are particularly interested in examining whether signed networks exhibit
different behaviours with regards to information dissemination.

In summary, we make the following contributions:
\begin{itemize}
	\item We measure the effect that individual, as well as combinations of multiple parameters have
	on the plain vanilla \emph{asynchronous push \& pull} rumour spreading protocol over a wide range of 
	values.

	\item We demonstrate that full rumour propagation is highly inefficient in social networks. However, the
	protocol's efficiency is exhibited when the purpose is to inform large subsets of the node population, e.g., 
	90\% to	97\%, which is often sufficient for voting and quorum-based systems.

	\item In contrast to prior work, we take a pragmatic, empirical approach.  Our study is 
    based on a set of ten diverse real-world, publicly available social network datasets and
	our network model accounts for link latencies and bandwidths as well as the concurrent processing of
	network events at nodes.
 
	\item We present an \emph{Enhanced push \& pull} protocol that is based on an intelligent selection of parameter
	values, such as our novel neighbor selection policy, and illustrate that it improves the total time required to
	inform 100\% of the nodes by up to 99.69\%.
\end{itemize}

%% file: background.tex
\section{Background}
\label{sect:backg}
Rumour spreading protocols are a series
of randomized protocols which were initially proposed for distributing updates
and ensuring eventual consistency amongst the sites of a replicated database~\cite{alan.demers}.
Their simplicity, robustness, and scalability properties have made them attractive for use  
in a number of other applications including multicast~\cite{birman1999}, 
distributed ranking~\cite{chiuso2011}, and others.

In the simplest case, where there is only one piece of information to propagate, rumour spreading protocols resemble
the random phone call model introduced in \cite{richard.karp}. Each of these protocols assumes
a start-up phase, where a piece of information (called update, rumour, or gossip) is injected
at an arbitrary node, known as the originator. These algorithms then proceed in a series of
synchronous communication rounds, based on the period of a globally accessible clock. 
In each round, nodes can be in one of the following states: 
\emph{Informed}: a node that knows the rumour and will spread it;
\emph{Uninformed}: a node that has not yet received the rumour and will ask for it; 
\emph{Removed}: an informed node who will refrain from spreading the rumour since it no
longer considers it ``hot'', i.e., it is old news. The purpose of the last state is to limit
the amount of redundant communication while still trying to achieve rumour dissemination
to all nodes.

There are three basic versions of rumour spreading protocols.  In the
\emph{push} version, only the informed nodes choose uniformly at random
a neighbor to which they will transmit
the rumour. This requires the sending of only one message which, when
sent, marks the end of the current round. In the \emph{pull} version,
every uninformed node contacts a randomly selected
neighbor and asks it for the rumour. The recipient of the \emph{pull} message replies by
sending back the rumour, \emph{iff} it is informed (if it is uninformed, the reply will be
an empty message). Note that in this version, two message exchanges are required for
a round to complete.

One can combine the two aforementioned strategies and obtain the \emph{push \& pull}
version, where, in each round, every node chooses uniformly at random one of its neighbors and,
depending on whether it is informed or not, either pushes or pulls the rumour. Therefore, an
informed node can, in a single round, push the rumour to one of its neighbors and also inform
one additional node if it also receives a \emph{pull} message from the latter.
Thus, in this setting, a round can involve the exchange of up to a total of three messages.
Nearly all prior publications (see Section~\ref{sect:relatedw}) on rumour spreading protocols
provide upper bounds regarding the number of rounds that are required for a rumour to spread
across all nodes.

To avoid the constraints of synchrony, researchers have proposed asynchronous variants
of these protocols in which every node is equipped with its own independent clock. These clocks 
comply with the asynchronous time model introduced by Boyd et al.~\cite{stephen.boyd}.
Namely, node clocks are modeled as rate 1 Poisson processes, i.e., the time between two consecutive
clock ticks is independent and exponentially distributed with $\lambda=1$.

%% file: relatedwork.tex
\section{Related Work}
\label{sect:relatedw}

The literature concerning rumour spreading protocols is extensive.
A number of studies focus on how the topology (i.e., graph structure) of connections between 
nodes affects the number of rounds required for a rumour to spread throughout the system for a variety of 
graph topologies including meshes and tori~\cite{Juurlink1996}, butterfly networks~\cite{Sibeyn2005},
sensor networks~\cite{stephen.boyd}, random, regular, Erd{\H{o}}s-R{\'{e}}nyi 
and social graphs (\cite{bj.doer2, flavio.chie0, flavio.chie1, flavio.chie2, bj.doer1, nikos.fountoulakis0, chrysis.georgiou}).  
Other studies propose constructing or imposing a particular structure on the topology of connections
between nodes in the system
to enable more efficient dissemination~\cite{kermarrec}.

Some studies focus only on the synchronous versions of randomized rumour spreading
(\cite{flavio.chie0}, \cite{flavio.chie1}, \cite{flavio.chie2},
\cite{bj.doer1}) whilst others (\cite{bj.doer0}, \cite{bj.doer2},
\cite{nikos.fountoulakis0}, \cite{chrysis.georgiou}) consider both synchronous and asynchronous.
Regarding the former set, it is well-known that achieving perfect synchrony in a distributed
system amongst thousands of nodes distributed over a network with heterogeneous
link latencies and bandwidths is difficult, at best.
Moreover, recent results show that asynchronous
rumour spreading protocols perform better in PA graphs than their
synchronous counterparts (\cite{bj.doer0}, \cite{bj.doer2}). For these reasons, we focus
on asynchronous rumour spreading in this work.

The majority of studies that focus on graph structure are theoretical analyses and aim
to determine the number of rounds that are required to inform all nodes. 
The authors start with a series of assumptions, typically regarding the graph structure,
and via a series of theorems and lemmas,
reach a conclusion that is along the lines of: ``In graphs with these properties,
under protocol $X$ and with high probability (w.h.p.), $O(x)$ rounds suffice
to broadcast a single piece of information to all nodes''. For instance, in
\cite{flavio.chie1}, the authors prove that the synchronous \emph{push \& pull}
protocol can, w.h.p., broadcast a message within $O( \frac{log^{4}{n}}{\phi^{6}} )$
rounds to all nodes. In a later publication (\cite{flavio.chie0}), this bound is
improved to $O( \frac{log^{2}{\phi}^{-1}}{\phi}  {logn} )$ for
PA graphs. These and other findings (\cite{bj.doer1}, \cite{nikos.fountoulakis0})
are important because they provide 
evidence that rumours may ``spread fast in social networks'' (quoting \cite{flavio.chie0}),
which is our focus of interest here.   However, they do not provide information
regarding network-related metrics.  Such information includes measured time delays and network load
incurred by the exchange of protocol messages, which can provide useful insight for the
design and implementation of practical systems. Moreover, these studies suggest that
nodes use derived formulas to
determine when to stop propagating a rumour to avoid sending redundant traffic over the network.
Unfortunately, these formulas are not feasible to implement in practice because 
they use values that are difficult to compute and/or require global graph knowledge such as graph
conductance or the total 
number of nodes in the system.  In social networks, where there may be tens of millions
of nodes, it is infeasible to have nodes compute, store, and update this type of global graph information
on a regular basis.  In our work, we take an empirical approach.  Our aim is to study 
how asynchronous rumour spreading behaves in \emph{real} social network settings 
and to find ways to design such protocols without the need for global graph knowledge.

To our surprise, despite the large number of prior works analyzing rumour spreading protocols, 
we find there is very little published work on optimizing or improving their performance.  Instead, the main
focus of the majority of works is the study of the plain vanilla versions of the push, pull, and push \& pull protocols.
There are three exceptions to this.  First, Georgiou et al.~\cite{chrysis.georgiou} examine fault
tolerance in asynchronous gossiping and propose algorithms that increase the robustness of the
protocol on randomized graphs in the face of an adaptive and oblivious adversary hampering rumour
dissemination. Second, Karp et al.~\cite{richard.karp}
attempt to minimize the amount of traffic generated by the asynchronous \emph{push \&
pull} protocol on random graphs by introducing the median-counter algorithm.  This is a
\emph{stopping criterion} nodes use to decide when to stop considering a
rumour ``hot'' and thus stop propagation. We examine the performance of this algorithm and others
in Section \ref{sect:experimentsandresults}.
Third, Doerr et al. \cite{bj.doer0} study the impact of equipping nodes with some ``neighbor memory'' that enables them
to avoid contacting the same neighbor twice in a row.
When the neighbor memory has space to hold the identity of one neighbor, Doerr et al.~\cite{bj.doer0} show that the performance
of the protocol improves by a factor of $\Theta(\log{\log{n}})$.  We note that this last study is the only work
of which we are aware 
that contains experimental results based on PA and real-world social
datasets. However, their results are for the synchronous version of the protocol. Here, we study
how the \emph{asynchronous} version of the protocol performs under varying sizes of neighbor memory
and on a more diverse set of topologies. 

In summary, while existing studies focus on isolated protocol parameters and their impact on 
performance, {\bf no} prior study has focused on how \emph{multiple} parameters affect the protocol's behaviour
both in {\emph{isolation and combination} and under a wide range of values. Such parameters include: choice of neighbor
from which to push and/or pull a rumour, the use of neighbor memory to aid in faster dissemination and reduce network load,
stopping criteria to end dissemination efficiently, and others. Our work is a first step in understanding how all of these
factors together affect protocol performance.

%% file: methodology.tex
\section{Methodology}
\label{sect:method}
Our simulation engine is a C++ adaptation of Narses (\cite{narses}), a Java 
discrete-event simulator, which we wrote from scratch.
Since there are no publicly available network traces (that
we know of) that indicate how social network nodes are connected at the IP layer,
we chose to model the underlying network topology as a star.
Nodes are linked to its center, which is assumed to have infinite
packet switching capacity, i.e., we assume that the core of the network has no
bottleneck links and that traffic is limited only by the end-link connections.
Social-link-based applications are comprised of peers that lie at the edges of 
the network which tend, in most cases, to be the most limiting factor (\cite{bottleneck}).
This approach, while simple, allows us to provide insight with
regards to network-related metrics, without sacrificing accuracy (\cite{narses}),
such as measured time delays and the network load incurred by the exchange of
messages.

Values for the links' bandwidths are chosen uniformly at random from the range
of [3,50] Mbps. These are all cheap, commodity and widely available DSL
line speeds. However, bandwidth speeds are not a significant factor in our experiments
since the messages that the nodes exchange are only a few
bytes long. Link latencies are also uniformly distributed from 10 to 100 ms
(\cite{link.pings1}, \cite{link.pings2}). 

\begin{table}[ht!]
	\centering
	{
		\begin{tabular}{l l c c c c c c}
			\\ \hline
			\textbf{Network} & \textbf{Type} & \textbf{n} & \textbf{e} 
			\\ \hline
			Slashdot1~\cite{slashdot1.dataset} & signed & 70,491 & 396,378 \\
			Slashdot2~\cite{slashdot2.dataset} & signed & 74,899 & 422,349 \\
			Slashdot3~\cite{slashdot3.dataset} & signed & 75,144 & 425,072 \\
			Epinions~\cite{epinions.dataset} & signed & 114,222 & 717,129 \\
			WikiSigned~\cite{wikisigned.dataset} & signed & 126,514 & 650,444 \\
			Hamsterster~\cite{hamsterster.dataset} & undirected & 1,858 & 12,533 \\
			Brightkite~\cite{brightkite.dataset} & undirected & 58,228 & 214,078 \\
			Facebook~\cite{facebook.dataset} & undirected & 63,731 & 1,545,686 \\
			TwitterLists~\cite{twitterlists.dataset} & directed & 23,370 & 33,101 \\
			Google+~\cite{googlep.dataset} & directed & 23,628 & 39,242 
		\end{tabular}
		\caption{Structural characteristics of the largest connected component (LCC) of the real-world social 
			 network datasets used in our simulations. \emph{Type}: connections
			 between users can be one-sided (directed), two-sided (undirected)
			 or even indicate a strong value of trust (signed); \emph{n}: number
			 of nodes; \emph{e}: number of edges.}
		\label{tablesocialstats}
	}
\end{table}
In our simulations, the ``rumour'' is simply a randomly selected long value. The
simulator, at time \emph{t} = 0, selects one node (typically this
is a popular node, but more on this in Section \ref{sect:experimentsandresults})
to play the role of the rumour's originator from which this information will start to spread.

The overlays, i.e., the social relationships amongst the users of the system, are
based on a large set of real-world, publicly available social graphs. Table
\ref{tablesocialstats} provides a listing, as well as some statistics, regarding
the largest connected component (LCC) of the datasets that we used in our simulations.

Nodes are initialized by receiving a list containing their social acquintances.
Each node then sets a timer that is programmed to
fire after an exponentially distributed time period with parameter $\lambda$. When a
node's timer expires, its behaviour depends on its current state: \emph{Informed}: The node
selects one or more neighbors, at random, and sends a \emph{Push} message; \emph{Uninformed}:
The node selects one or more neighbors, at random, and sends a \emph{Pull} message;
\emph{Removed}: The node does not schedule any more timers since it considers the rumour
that it is currently circulating in the network ``old news''.

Since we are interested in determining how quickly a piece of information can spread over
social graphs, we assume that nodes do not perform any computation before they send or
receive a message.
We use two metrics to evaluate protocol behaviour in our experiments.  First,
we measure the total time that is required to
inform either all, or, specific percentages of each graph's LCC.
Second, because rumour mongering protocols are known to produce large amounts of 
network traffic (\cite{kermarrec}), we also measure the load imposed on the network.
Simply plotting the number of produced messages
that are required to achieve some percentage of informed nodes over time does not suffice
because the graphs of the social networks we study vary in size significantly.  The protocol must
produce more messages over a larger period of time to achieve the same target percentage for larger
graphs compared to smaller ones.  Thus, the network load metric must be independent of the number of 
nodes in the graph to allow us to evaluate protocol behaviour in a manner that is consistent
across all social graphs.  We achieve this by plotting the network load 
as the ratio of the number of generated messages
to the total time needed to achieve the desired percentage of informed nodes.


%% file: experimentsandresults.tex
\section{Experiment Description and Results}
\label{sect:experimentsandresults}
In this section we present the
experiments we ran to evaluate the effect of different parameters on
the performance of the asynchronous \emph{push \& pull} protocol. 
\subsection{Single Parameter Experiments}
\label{subsect:singleparamexp}
{\bf Performance bottleneck.} The first thing that we observed when we began to simulate
this protocol was the large amount of time that is required to inform all of the nodes
in the LCC of each graph. For instance, in the Google+ topology, whose LCC is comprised
of 23,613 nodes, and with nodes engaging in the protocol at exponentially distributed
intervals with a mean value of 1 second, we find that it takes 20,238.75 seconds for all
the nodes to become informed, on average. This means that nodes become informed with a
rate of approximately 1.16 nodes/second, which is quite slow.

\begin{figure}[ht!]
  \centering
  {
      \includegraphics[width=0.45\textwidth]{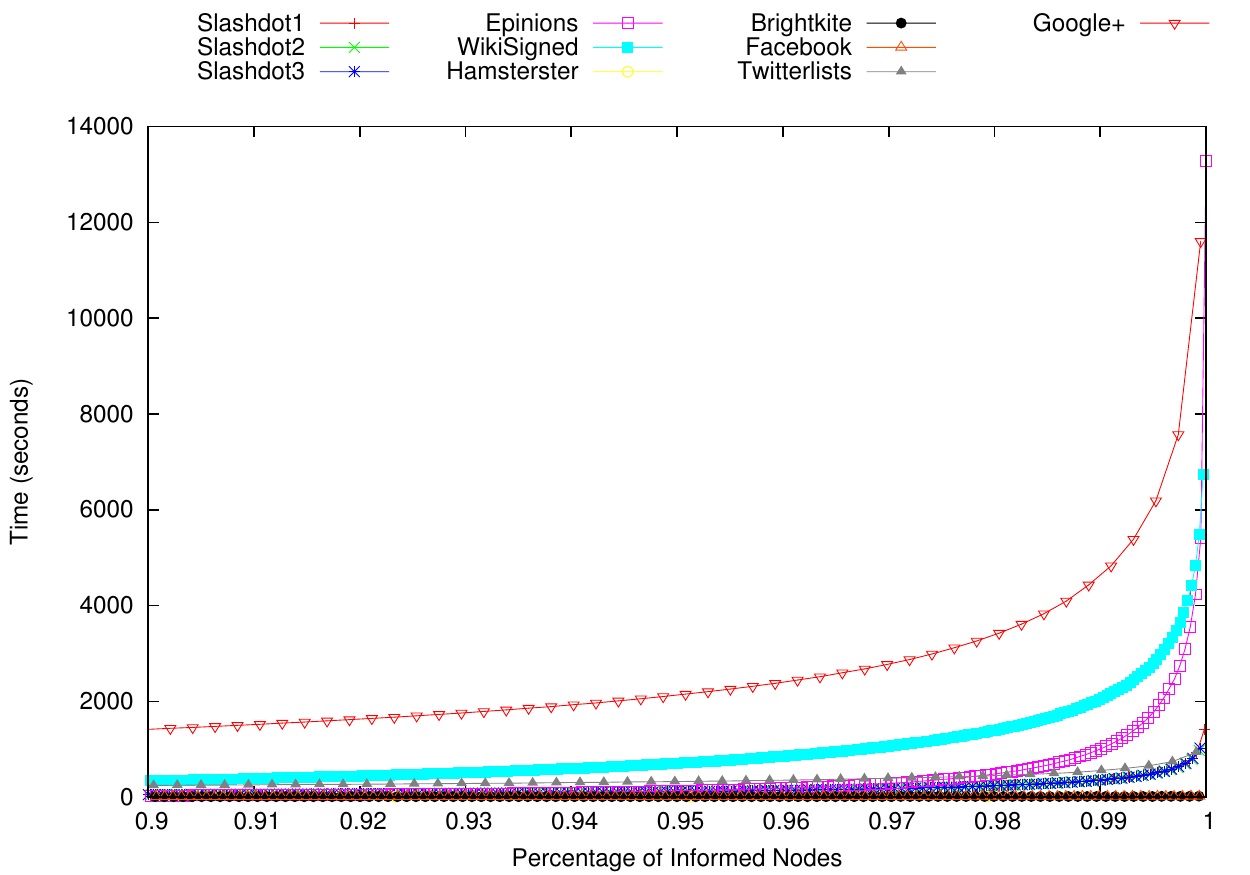}
  }
  \caption{Time required to inform the remaining 10\% of the LCC of each graph.}
\label{figure:exp8fig}
\end{figure} 
To understand why, we plot the total amount of time that is required to inform specific
percentages of the nodes in the LCC of each graph. 
Figure \ref{figure:exp8fig} shows how the time needed to inform the last 10\% of the LCC
of each graph scales.
Starting up from 90\% and all the way up to (and including) 96\% of
informed nodes, time scales smoothly (as is the case in lower percentages that are not
included in this plot). However, we observe a knee in the graph at 97\% after which the
total time increases exponentially. Through extensive analysis, we find that this extreme
delay is caused by some communities which are connected to the core of the network
via just a handful of links. Since the number of available paths that can be used to reach the
members of these communities is extremely small, their chances of being randomly selected
for information dissemination decreases. Furthermore, in some other cases, we also observe
that these communities are also connected, again via a handful of links, to other communities
where, the latter, are almost never connected to the core of the network. Thus, chains of
small communities are formed. For instance, assume that community A is connected via a
couple of links to the core of the network and is also connected to another
community, B, again via a handful of links. If community B is not directly linked to the
core of the network, then its members will become eligible for receiving information only
when the ``hub nodes'' of community A (the nodes connecting communities A and B) become informed.
This creates a convoy phenomenon, which makes it  harder for the protocol to inform these nodes
and thus the increased delay.

\begin{table}[ht!]
  \centering
  {
	\begin{tabular}{l c c c}
	\hline
	\textbf{Network} & \textbf{97\% Inf.} & \textbf{100\% Inf.} & \textbf{\% Incr.}
	\\ \hline
		Slashdot1 & 184.27 & 2,470.17 & 1,240.46\% \\
		Slashdot2 & 179.93 & 2,572.21 & 1,329.53\% \\
		Slashdot3 & 179.26 & 2,438.73 & 1,260.38\% \\
		Epinions & 289.19 & 13,289.52 & 4,495.32\% \\
		WikiSigned & 1,064.83 & 15,308.48 & 1,337.64\% \\
		Hamsterster & 9.10 & 21.33 & 134.38\% \\
		Brightkite & 12.65 & 32.54 & 157.14\% \\
		Facebook & 12.28 & 28.40 & 131.24\% \\
		TwitterLists & 394.56 & 1,588.44 & 302.58\% \\
		Google+ & 2,780.38 & 20,238.75 & 627.91\%
	\end{tabular}
  }
	\caption{Total time required to inform 97\% and 100\% of the nodes in the largest
	connected component of each graph and the percentage increase of the latter case
	compared to the former.}
	\label{table:experiment8stats}
\end{table}
Table \ref{table:experiment8stats} illustrates the total amount
of time required to inform 97\% and 100\% of the nodes in the LCC of each graph, as well
as the percentage increase of the latter compared
to the former. From Table \ref{table:experiment8stats}, we see that the protocol exhibits 
significant difficulty informing these last few node percentages, especially in signed
and directed topologies. Bidirectional topologies suffer a much smaller decline. This is
attributed to the way that information flows through
the chained communities that we mentioned in the previous paragraph. In signed and directed topologies, in order for the rumour to
spread throughout the chain, it either needs to be pushed or pulled from one hub node
to the next, since information flows only one way. Keep in mind that whether the next
hub node is selected or not is still a random choice. In most cases the protocol
will select one of the few community members instead, which throttles the rumour's propagation
even more. In bidirectional topologies, the rumour can be both pushed and pulled as
it travels throughout the chain, since information flows both ways. In addition,
the members of these chained communities can become informed much faster since they have
the added ability of pulling the rumour from their community's hub node moments after it
becomes informed. The fact that the protocol can exploit the benefits
of both push and pull results in a significant reduction of the rumour's spreading time
throughout the chain in bidirectional topologies. To our knowledge, ours is the first study
to document this convoy phenomenon.  This is important because many applications (e.g., voting
or quorum-based systems) may not need full propagation to all nodes to make progress.
Since this phenomenon also affects the readability of graphs that plot time over node percentages,
from hereon in, for most of the graphs, we plot data values whose maximum node percentages lie 
in the range of [90\%, 97\%].

{\bf Originator popularity.} We next study the effect that the connectivity, or, 
``popularity'' of a node has on the dissemination of a rumour.
We classify the nodes of the social graph into three different groups inspired by the scheme
that is presented in \cite{kumar}. Nodes reside in one of the following categories according
to their out-degree value:
\begin{itemize}
	\item Giant Component (Group 3): This group consists of highly connected individuals
	that are connected with a large fraction of the network. For instance, singers, actors,
	or other famous individuals are typical members of this category. 
	\item Middle Region (Group 2): These are star-like shaped communities, with one or two nodes at
	the center, that mostly interact with other fellow group members and are sparsely connected
	to the giant component.
	\item Singletons (Group 1): These are the one-degree nodes.
\end{itemize}

\begin{figure}[ht!]
 \centering
 {
    \includegraphics[width=0.45\textwidth]{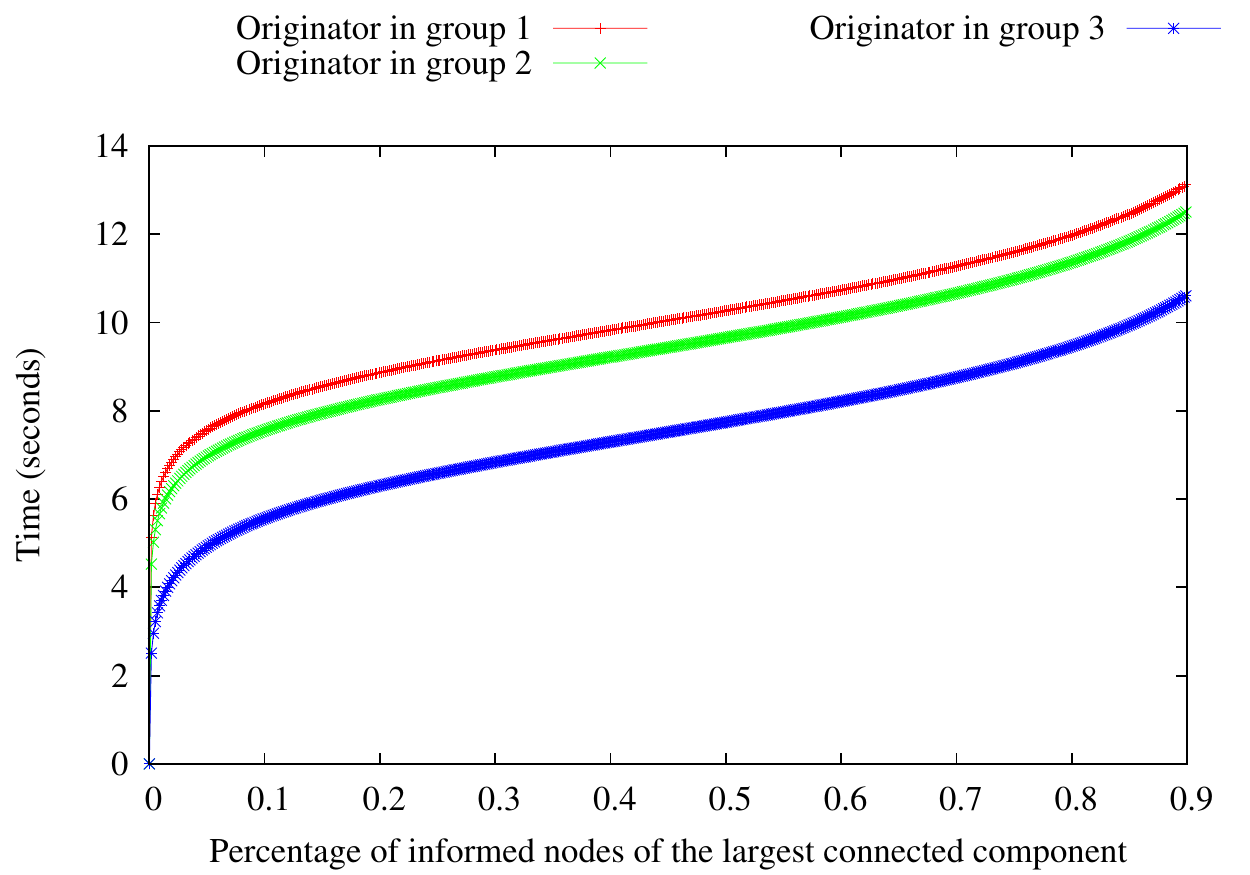}
 }
 \caption{Effect of the rumour's originator group membership on the protocol's performance for
the Brightkite topology.}
\label{figureexp2fig}
\end{figure}

For each of the aforementioned groups, we randomly sample 10\% of its members to act as
originators of rumours and measure the impact this has on the performance of the protocol. We
repeat this procedure 1,000 times for each distinct source and average the results. In  Figure
\ref{figureexp2fig},
we plot the effect of originator
popularity on the time to inform varying percentages of nodes for the
Brightkite topology. This is a representative plot we have chosen due to lack of space. Other
datasets exhibited similar results.

The results illustrate an important difference in the total time required to inform 90\%
of the nodes in the LCC of the Brightkite topology. The data suggest an average slowdown of
4.69\% and 19.14\%, when the originator is in group 1, compared to the case where the originator
is in group 2 and 3, respectively. The best result is obtained in the Epinions topology, where a
popular originator (group 3) can improve the total time required to inform 90\% of the nodes by an
average of 28.46\%. Across all topologies, and for the same percentage of informed nodes, the average slowdown is
10.96\% and 15.71\%. 

{\bf Neighbor Contact Criteria.} The fact that a node's popularity can speed up
the dissemination of a piece of data, inspired us to investigate if rumour dissemination 
can be accelerated if nodes picked their communication partners based on a scheme that
favors their most popular, or well-connected, neighbors. This biased selection deviates from the standard
uniform random choice used by almost all prior works. We explore the effects of the following
alternative neighbor contact criteria:
\begin{itemize}	
	\item Quasirandom (Q): Nodes have a cyclic list of friends. They choose a random starting position and from
	then on, they contact neighbors in a round robin fashion. This model is presented in \cite{bj.doer3}.
	\item Quasirandom Popular (QP): Nodes sort their lists according to popularity and cycle through
	them from the most popular to the least popular node.
	\item Quasirandom Unpopular (QU): The exact opposite of QP, i.e., nodes cycle from the least
	popular to the most popular node.
	\item Quasirandom Popular to Unpopular (QPU): Neighbor lists are again sorted based on neighbor
	popularity and nodes first choose the most popular node, then the least popular, then the second
	most popular, then the second least popular etc.
	\item Quasirandom Unpopular to Popular (QUP): The exact opposite of QPU.
\end{itemize}

In Table \ref{table:neighborstats}, we illustrate the performance improvements of these strategies
compared to the standard uniform random choice. We note that these are all novel neighbor contact
criteria that have not been evaluated in the past (apart from quasirandom).

\begin{table}[ht!]
  \centering
  {
	\begin{tabular}{ >{\centering\arraybackslash}m{1.4cm} >{\centering\arraybackslash}m{0.8cm} >{\centering\arraybackslash}m{0.8cm} >{\centering\arraybackslash}m{0.8cm} >{\centering\arraybackslash}m{0.8cm} >{\centering\arraybackslash}m{0.8cm} }
	\\ \hline
	\textbf{Network} & \textbf{Q} & \textbf{QP} & \textbf{QU} & \textbf{QPU} & \textbf{QUP}
	\\ \hline
		Slashdot1 & 82.75\% & 82.97\% & 82.67\% & 82.73\% & 82.64\% \\
		Slashdot2 & 82.43\% & 82.55\% & 82.38\% & 82.37\% & 82.44\% \\
		Slashdot3 & 82.65\% & 82.93\% & 82.61\% & 82.60\% & 82.56\% \\
		Epinions & 84.70\% & 84.80\% & 84.82\% & 84.86\% & 84.90\% \\
		WikiSigned & 82.57\% & 82.89\% & 83.73\% & 82.82\% & 83.79\% \\
		Hamsterster & 10.82\% & 21.79\% & 2.44\% & 21.44\% & 15.39\% \\
		Brightkite & 32.88\% & 37.11\% & 27.26\% & 36.70\% & 36.83\% \\
		Facebook & 27.06\% & 34.82\% & 21.11\% & 33.21\% & 32.07\% \\
		TwitterLists & 80.29\% & 78.54\% & 78.53\% & 80.03\% & 79.50\% \\
		Google+ & 85.12\% & 85.13\% & 84.88\% & 84.93\% & 85.17\%
		\\ \hline \hline		
		Average & 65.13\% & 67.35\% & 63.04\% & 67.17\% & 66.53\%
	\end{tabular}
  }
	\caption{Percentage improvement of the total time required to inform 100\% of the nodes
	in the LCC of each graph that each alternative neighbor selection strategy delivered.}
	\label{table:neighborstats}
\end{table}
All of the variants of the quasirandom model provide tremendous improvements. The data in
Table \ref{table:neighborstats} suggest that the best options are QP, QPU and QUP (their performance is almost tied) since they
strike a nice balance between using popular nodes and their many links to disseminate the rumour,
and unpopular nodes, who are the major cause of delay, as we illustrated in the beginning of this
section. However, in our combined experiments, we find that QPU performs better
when combined with other protocol parameters.  

{\bf Neighbor memory.} The use of ``neighbor memory'', i.e., the ability of nodes to remember
with which nodes they have communicated and to avoid communicating with them in the future,
has been shown to  improve the performance of the synchronous \emph{push \& pull} protocol,
even when it has space to hold only a single neighbor (\cite{bj.doer0},
\cite{bj.doer2}). For instance, Doerr et al. \cite{bj.doer0} illustrate that equipping the
synchronous \emph{push \& pull} protocol with a 1-item memory yields an improvement of about
14\%-21\%, compared to the case where there is no memory, in PA graphs. 

To date, no prior publication has examined how the \emph{asynchronous}
version of the \emph{push \& pull} protocol performs under varying neighbor memory sizes $m$.
We explore how the use of neighbor memory can enhance the performance of the ``Random'' neighbor
selection strategy by varying $m$ from 1 to 10. Note that, in contrast to the quasirandom variants,
the ``Random'' neighbor selection strategy is the only one that does not use any memory by default. We find performance improvements only in 
bidirectional topologies. 
\begin{table}[ht!]
  \centering
  {
	\begin{tabular}{c | l | c | c | c}
		\\ \hline
		\textbf{Memory Size} & \textbf{Network} & \textbf{90\% Inf.} & \textbf{99\% Inf.} & \textbf{100\% Inf.}
		\\ \hline
		\multirow{3}{*}{1 vs 0} & Hamsterster & 1.35\% & 6.71\% & 16.55\% \\
		& Brightkite & 5.37\% & 8.97\% & 12.43\% \\
		& Facebook & 3.01\% & 4.96\% & 10.88\% \\
		\hline
		\multirow{3}{*}{2 vs 1} & Hamsterster & 2.56\% & 4.31\% & 1.46\% \\
		& Brightkite & 2.73\% & 3.69\% & 8.69\% \\
		& Facebook & 1.09\% & 1.66\% & 4.13\% \\
		\hline
		\multirow{3}{*}{3 vs 2} & Hamsterster & 3.65\% & 2.59\% & 1.56\% \\
		& Brightkite & 2.94\% & 5.24\% & 5.33\% \\
		& Facebook & 2.09\% & 3.13\% & 12.48\% \\		
	\end{tabular}
  }
	\caption{Comparative data regarding the percentage improvement of the total time required by the ``Random'' neighbor selection
	policy to inform 90\%, 99\% and 100\% of
	the nodes of all bidirectional networks for various neighbor memory sizes.}
	\label{table:nmrandomtimetable}
\end{table}
Signed and directed topologies do not benefit from varying this
parameter. We illustrate the percentage improvements that a 1, 2 and 3 item neighbor memory achieves
in bidirectional topologies, regarding the total time required to inform various node percentages,
in Table \ref{table:nmrandomtimetable}.
Note that the results in Table \ref{table:nmrandomtimetable}
are comparative, i.e., we illustrate the performance improvement that a 1-item memory yields compared
to the case where there is no memory, the performance improvement that a 2-item memory yields compared
to the case of a 1-item memory and so on.

In \cite{bj.doer0}, the authors illustrate that a 1-item memory reduces the total time required by
the \emph{synchronous push \& pull} protocol to inform 90\%, 99\% and 100\% of the nodes of
Orkut, a large, real-world, bidirectional topology, by 1.4\%, 1.6\% and 9.3\% respectively. Further
increases in the memory's size do not yield any significant improvements compared to the case of a 1-item memory.
Our results illustrate that, in bidirectional topologies, the asynchronous version of the protocol
benefits much more for the same informed node percentages. Furthermore, it seems that the asynchronous
version of the protocol can benefit from larger neighbor memory sizes. The data suggest a soft cap
at $m=3$; passing that point yields no benefits.
As the data in the table suggest, this simple enhancement can provide radical improvements
to the protocol's performance. These are especially evident when the objective is to inform
100\% of the nodes in the LCC. For instance, in the Facebook topology, a 3-item neighbor memory
can reduce the total time required to inform 100\% of the nodes by an average of 27.49\%.
\subsection{Combined Experiments}
\label{subsect:combexp}
Here, we attempt to leverage the performance benefits offered by the combination of
multiple protocol parameters. Our goal is to derive a practical protocol that manages
to disseminate a rumour to a large percentage of nodes, whilst imposing a minimal load on the network.

{\bf Stopping Criteria.} In a realistic setting, nodes cannot propagate a rumour indefinitely. Several theoretical
papers (\cite{flavio.chie0}, \cite{flavio.chie1}, \cite{bj.doer1}, \cite{nikos.fountoulakis0})
have proposed upper-bounds for the number of rounds that are required to inform either all,
or at least a large percentage of the nodes, w.h.p. These ``stopping criteria'' serve as a
means to limit the amount of redundant information that is circulating over the network.
We evaluate the performance of the following stopping criteria:
\begin{itemize}
	\item Several upper bounds such as: $\log_3{n} + O(\ln{\ln{n}})$  (\cite{richard.karp}), $O(n\log{n})$ (\cite{richard.karp}),
	$O(\log{n})$ (\cite{bj.doer1}), and $O(\log^{2}{n})$ (\cite{flavio.chie2}).
	\item Median Counter: This is a distributed termination algorithm proposed in \cite{richard.karp} for
	the synchronous \emph{push \& pull} protocol. We omit the full details here, but essentially, each node
	holds state (a counter) that functions as a way to track the progress of the rumour's propagation amongst
	its neighbors. It attempts to limit the number of times that a node pushes the rumour when it detects that
	its neighborhood is highly informed, with respect to the theoretical upper bound that the authors prove
	that is required. We implement an asynchronous version of this algorithm where nodes follow the rules of the algorithm
	based on their local perception of what a round really is.
\end{itemize}

\begin{figure*}[ht!]
	\centering
	{
	    \subfloat[]
	    {
	      \includegraphics[width=0.32\textwidth]{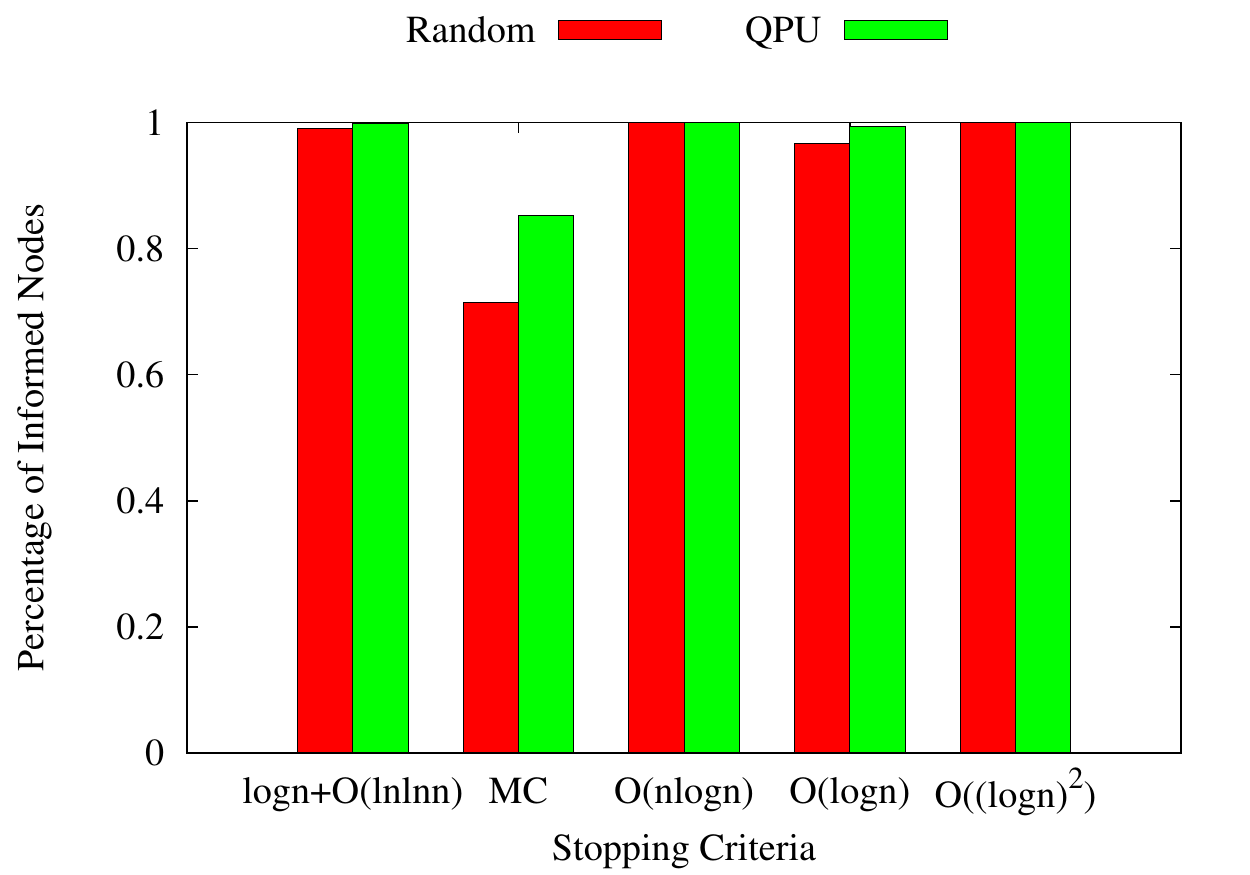}
	      \label{fig:neighborandstop1}
	    }
	    \subfloat[]
	    {
	      \includegraphics[width=0.32\textwidth]{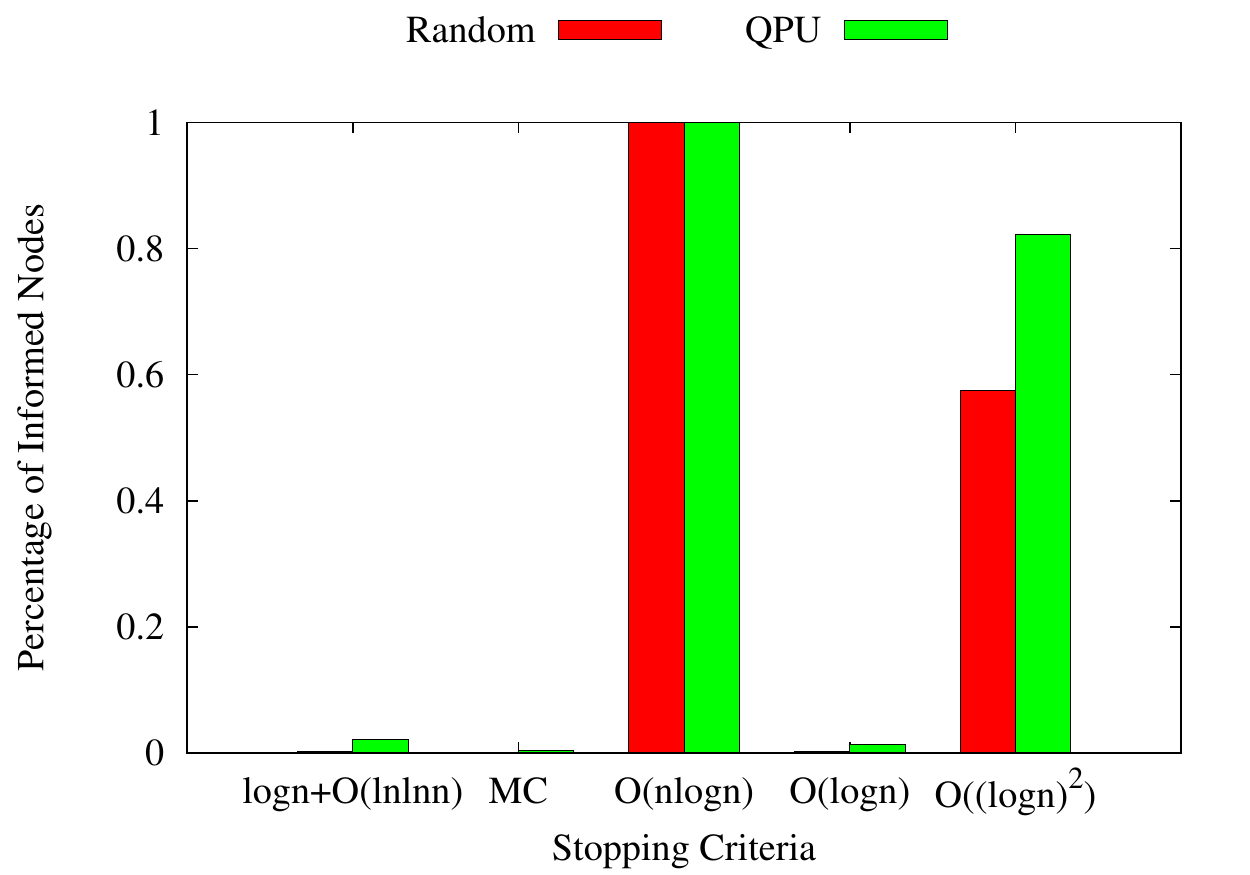}
	      \label{fig:neighborandstop2}      
	    }
	    \subfloat[]
	    {
	      \includegraphics[width=0.32\textwidth]{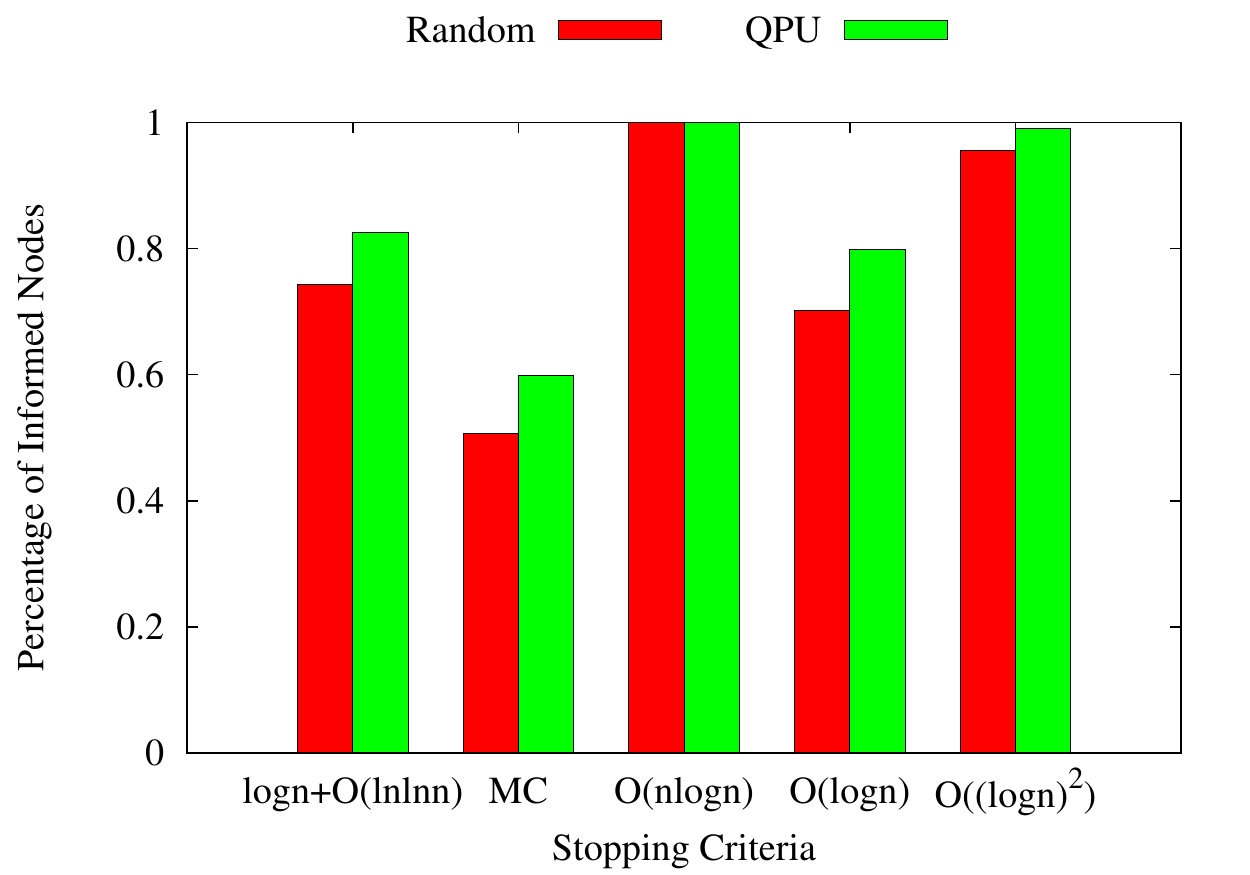}
	      \label{fig:neighborandstop3}
	    } 	        
	}
\caption{Average percentage of informed nodes on the Facebook (a), Twitterlists (b) and Slashdot2 (c) that the QPU and the random
neighbor contact scheme achieve when combined with all of the stopping criteria. MC stands for Median Counter.}
\label{fig:neighborandstop}
\end{figure*}

Our first approach is to combine the aforementioned stopping policies with a more efficient
neighbor contact scheme.  We experimented with all of the different neighbor contact criteria that
we previously presented
and found that QPU manages to deliver the biggest improvement (the next in line are QUP and QP).
In Figure \ref{fig:neighborandstop}, we present bar charts that illustrate the improvement in
the average percentage of informed nodes, for each stopping criteria, that this parameter combination
delivers, compared to the standard ``Random'' neighbor choice, in three different topologies.

According to the data, the most reliable stopping criteria is $O(n\log{n})$ since
it manages to inform all nodes across all topologies. While several theoretical papers (e.g. \cite{flavio.chie0}) have
illustrated in the past that $O(n\log{n})$ rounds are always sufficient for synchronous rumour spreading protocols,
we experimented with this upper bound for the following reasons. First, the behavior of synchronous and
asynchronous rumour spreading protocols in social networks is quite different. This is illustrated both in prior
work (e.g. \cite{bj.doer0}, \cite{bj.doer2}), as well as in our experiment section regarding neighbor memory.
Second, ours is the first study that evaluates the performance of the asynchronous \emph{push \& pull}
protocol when coupled with various theoretical upper bounds. Thus, we complement prior theoretical work and shed
light on how the protocol behaves in practical, real-world systems. 

While $O(n\log{n})$ manages to deliver perfect
performance, we advise against its use due to its insurmountable
cost. The following example will help illustrate this point; the Facebook topology, has a total of $n=63,392$ nodes, meaning
that the rumour's originator will have to push the rumour about 304,410 times, despite the fact that the most popular node
has an out-degree of just 1,098 (which is about 277 times less). Note, that these are the messages generated only by the
rumour's originator, i.e., one node. The second node that will receive the rumour will generate 1 message less than the
originator, the third node 2 messages less and so on. This will quickly cause an immense number of messages to be generated
that will impose severe load on the network. Thus, one should aim for more efficient approaches.

The introduction of the QPU neighbor contact scheme increases the achieved percentage of informed nodes by an
average of 11.59\% across all topologies. In all of the bidirectional
topologies, the $O(\log^2{n})$, $\log_3{n} + O(\ln{\ln{n}})$ and $O(\log{n})$ stopping policies combined with QPU achieve
informed node percentages that lie in the range of 98.68\% to 100\%, i.e., they perform well. In signed topologies, only $O(\log^2{n})$ manages to
deliver an average informed node percentage of 96.46\%. The performance of $O(\log{n})$ and $\log_3{n} + O(\ln{\ln{n}})$
degrades to an average of 76.53\% and 79.11\% respectively, in these topologies. In directed topologies, however,
only $O(\log^2{n})$ delivers an acceptable and high informed node percentage (82.23\% on Twitterlists),
thus making it the only viable option across all graphs.

The reason why most of the stopping criteria perform so poorly in these directed topologies is because
they have extremely low clustering coefficients and algebraic connectivities
(0.00371, 0.01143 for Google+ and 0.00215, 0.00503 for Twitterlists). These measures
indicate graph density and are generally known to have much
higher values in social networks. For instance, the Facebook topology has a clustering coefficient
of 0.148 and an algebraic connectivity of 0.51, which are almost two orders of magnitude larger than
the corresponding coefficients of Google+ and Twitterlists. Rumours spread well in social networks due to their 
high density, i.e.,
there are many links that the protocol can use to disseminate information to many different nodes.
Obviously, this does not hold for these directed topologies.

{\bf Communication Fan-Out.} We have thus far assumed that the communication fan-out parameter, i.e., the number $f$ of
neighbors with which a node communicates on each round, has a value of 1. However, there are variants
of rumour mongering algorithms (e.g. \cite{kermarrec}) that use larger values to provide greater
fault-tolerance in the presence of link-failures, or simply to disseminate information faster. We investigate
its effects on the protocol's performance by combining it with the QPU neighbor contact scheme and the
$O(\log^2{n})$ stopping policy under two different approaches. In our first approach, called absolute communication 
fan-out, we set a value $f_{abs}$ (for instance $f_{abs}=3$) which specifies the number of 
nodes with which every node will attempt to communicate with (it is possible that
some nodes will communicate with less than $f_{abs}$ neighbors because they do not have that many neighbors). We
vary the value of $f_{abs}$ starting from 1 to 15, in increments of one. In the second approach, called
relative communication fan-out, we fix a value $f_{rel}$
in the range of [0,1] which indicates the fraction of neighbors with which nodes will communicate
in every round. This means that nodes with larger neighbor lists will communicate with more nodes than
the ones with smaller lists. Nodes whose neighbor lists are too small, i.e., multiplying their lists sizes by
the fraction yields a value that is less than 1, are set to communicate with one of their neighbors. We
start by setting $f_{rel}=1\%$ and proceed in increments of 3\%, up to a maximum value of 25\%.

\begin{figure*}[ht!]
	\centering
	{
	    \subfloat[]
	    {
	      \includegraphics[width=0.32\textwidth]{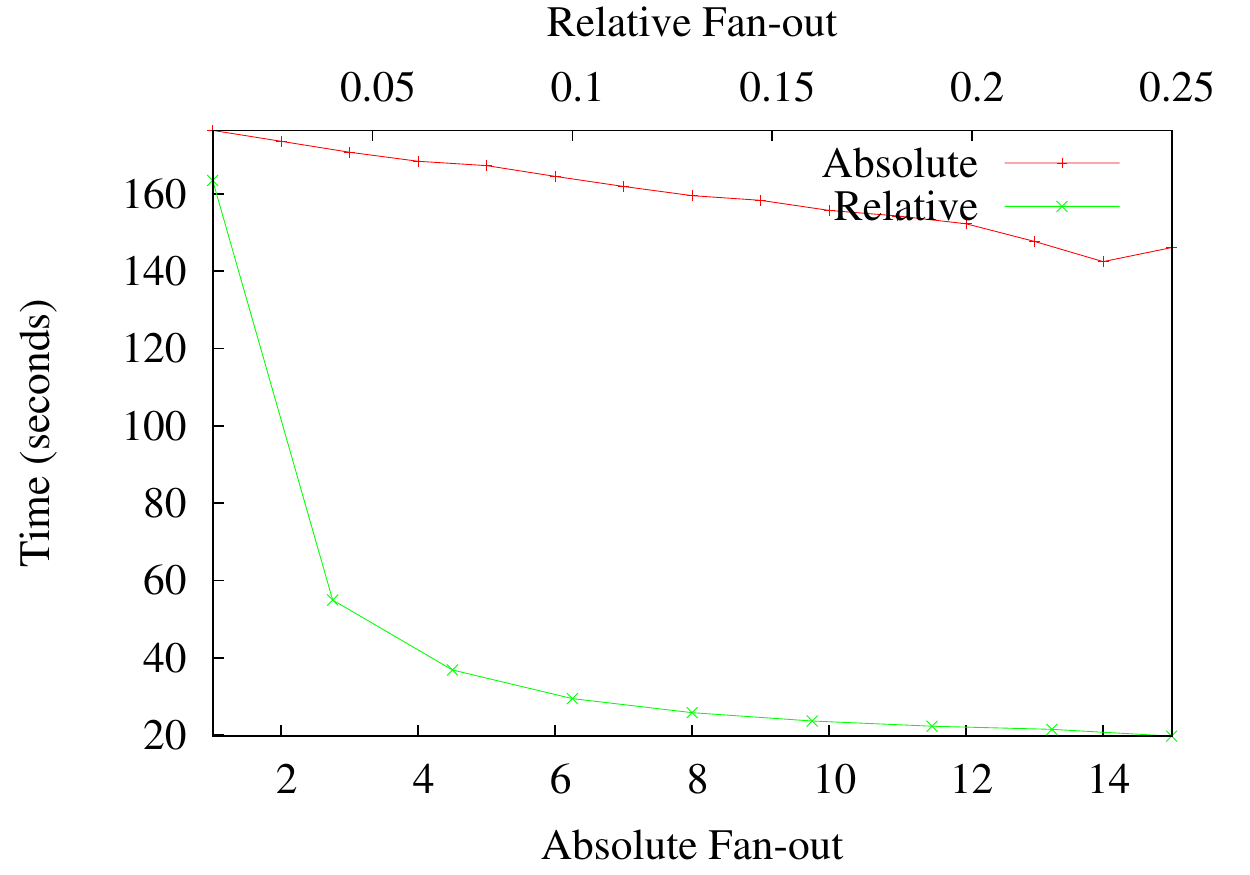}
	      \label{fig:neighborandstopandfanout1}
	    }
	    \subfloat[]
	    {
	      \includegraphics[width=0.32\textwidth]{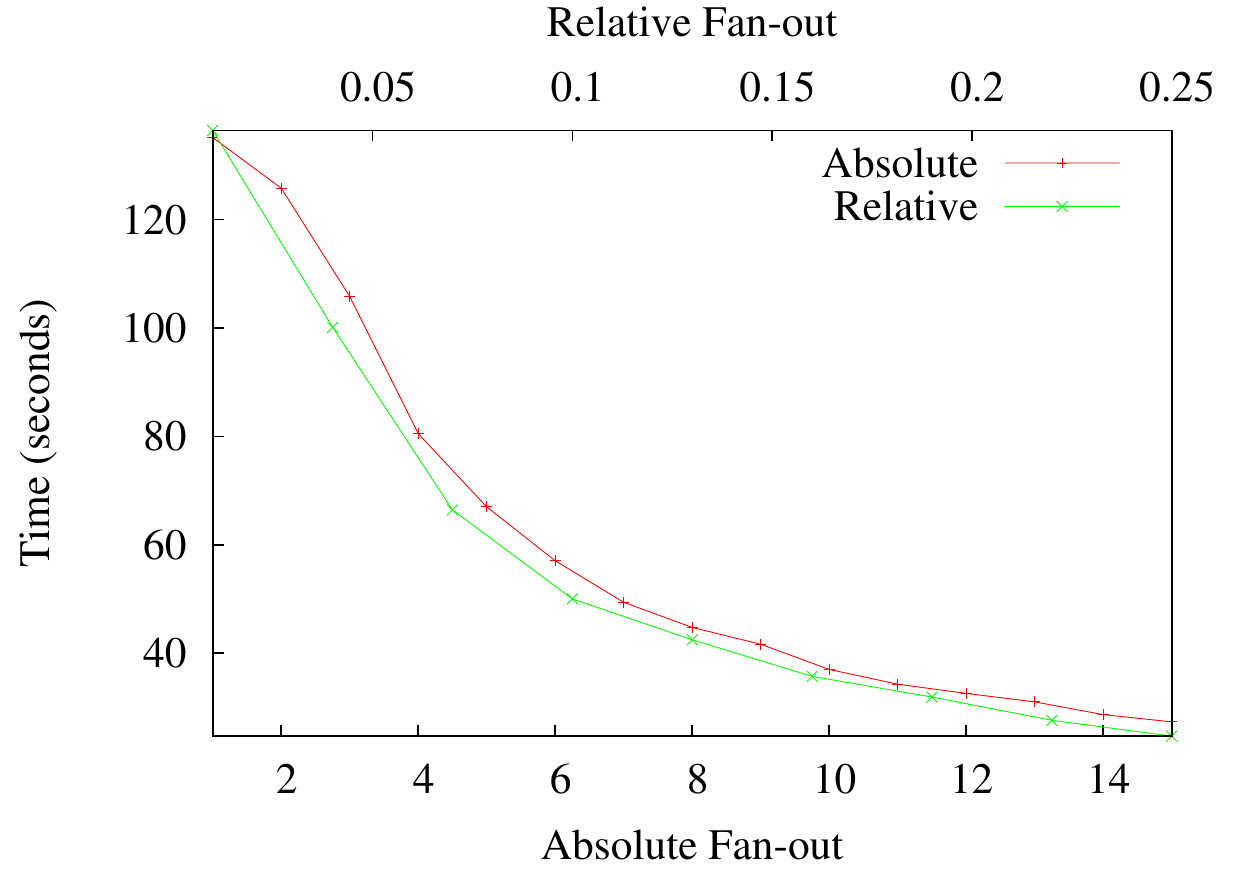}
	      \label{fig:neighborandstopandfanout2}      
	    }
	    \subfloat[]
	    {
	      \includegraphics[width=0.32\textwidth]{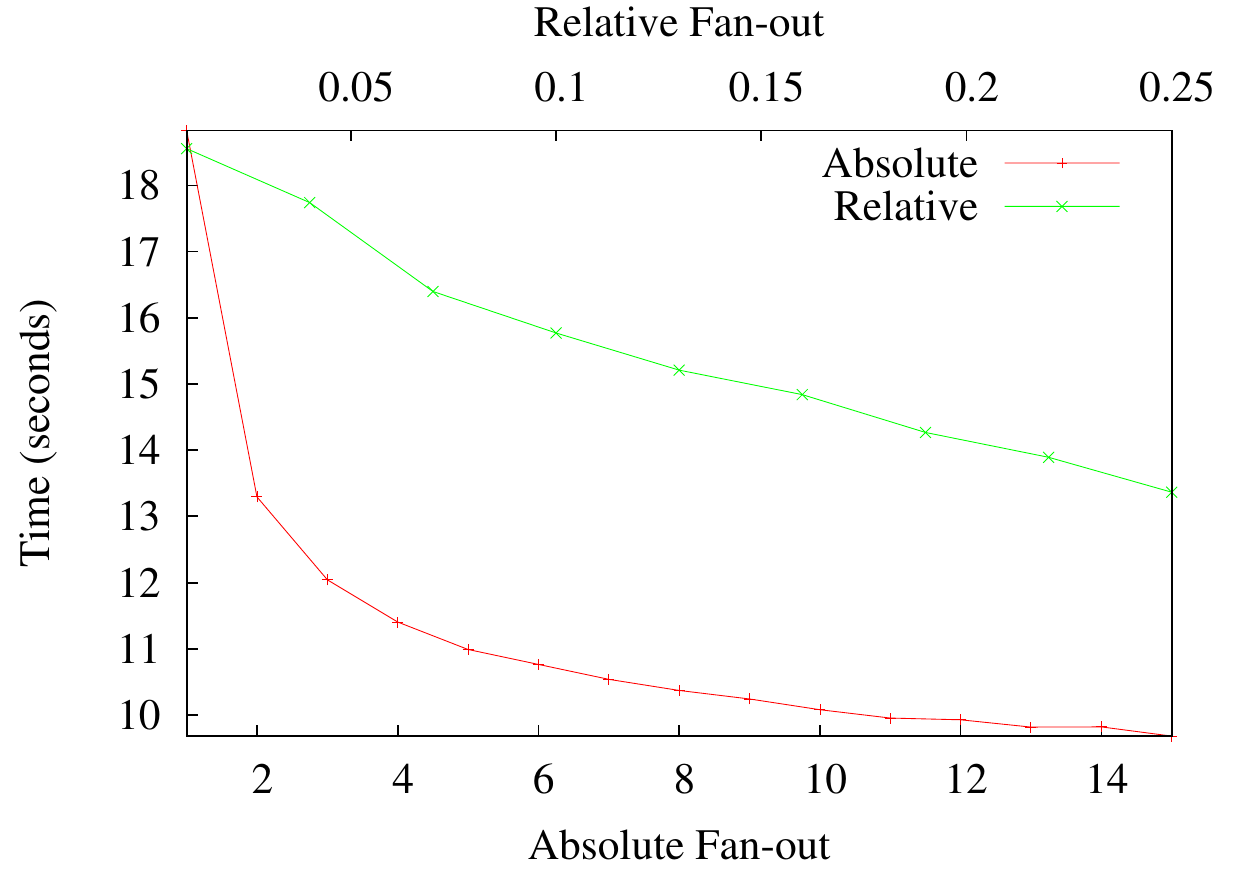}
	      \label{fig:neighborandstopandfanout3}
	    } 	        
	    \hfil
	    \subfloat[]
	    {
	      \includegraphics[width=0.32\textwidth]{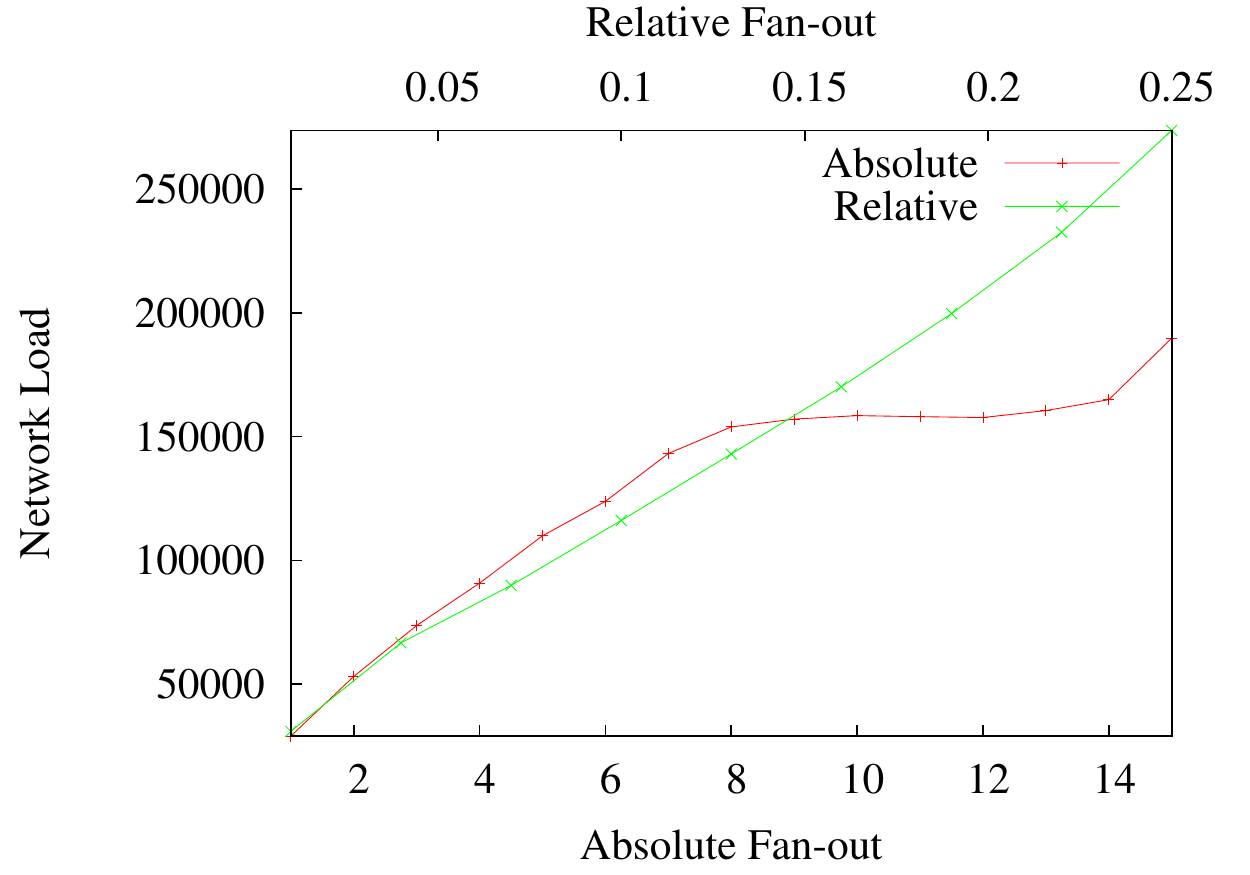}
	      \label{fig:neighborandstopandfanout4}
	    }
	    \subfloat[]
	    {
	      \includegraphics[width=0.32\textwidth]{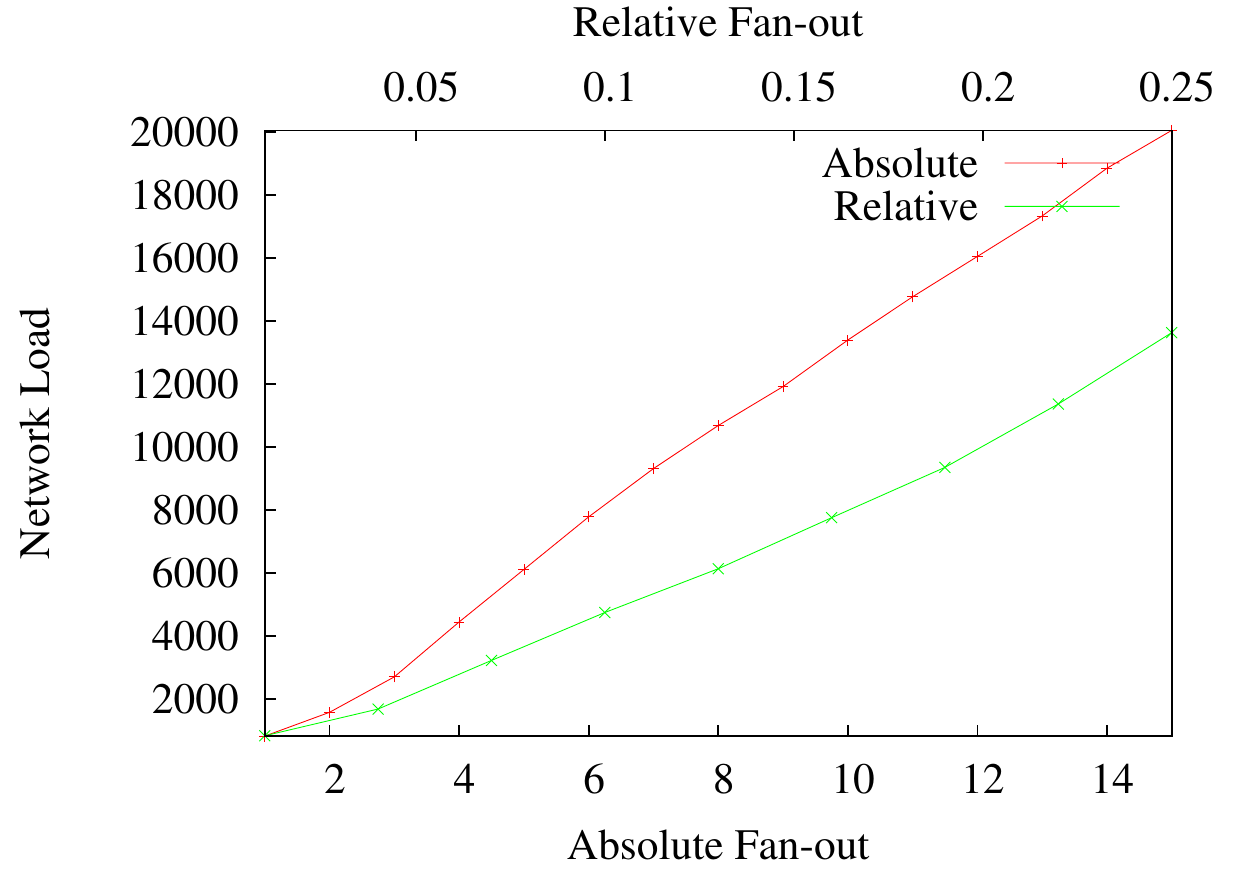}
	      \label{fig:neighborandstopandfanout5}      
	    }
	    \subfloat[]
	    {
	      \includegraphics[width=0.32\textwidth]{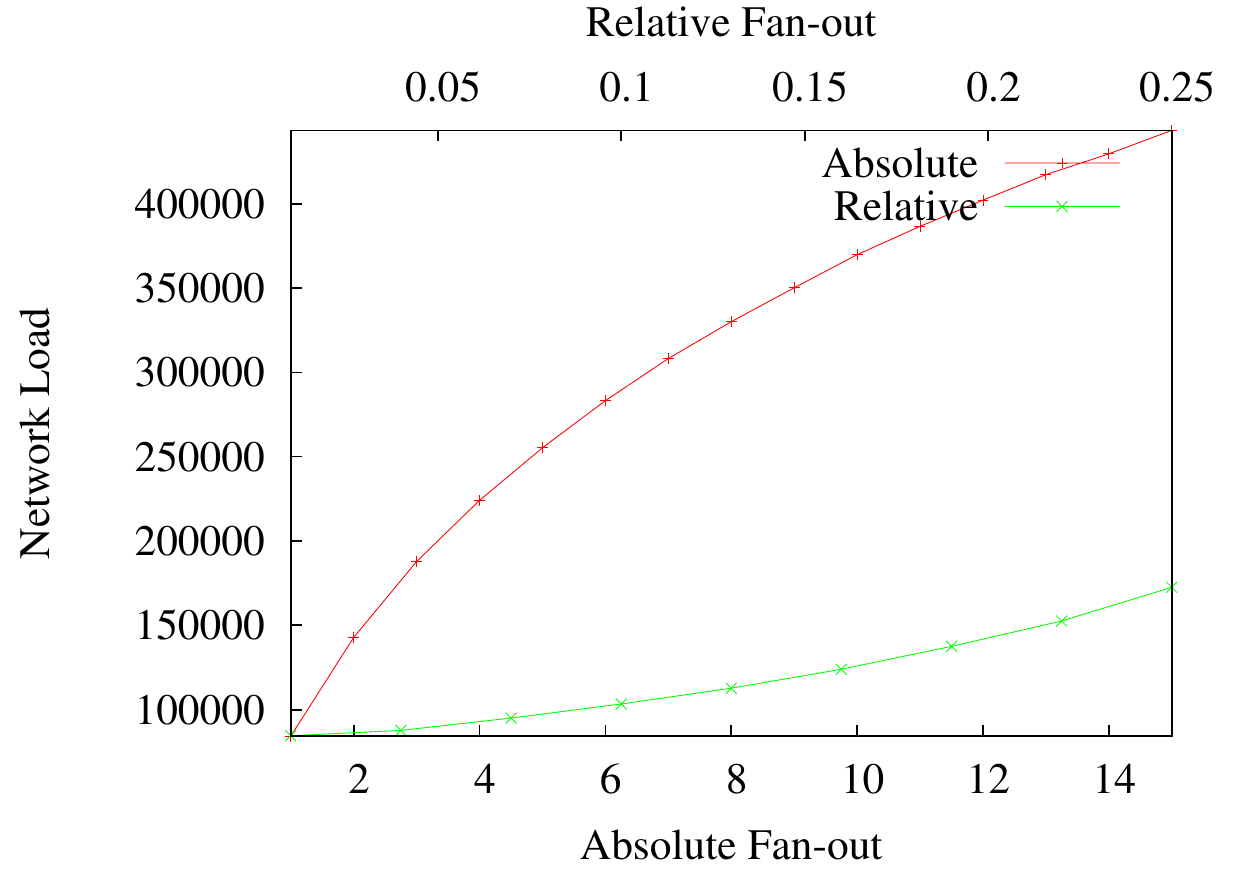}
	      \label{fig:neighborandstopandfanout6}
	    } 	    
	}
\caption{Starting from left to right, communication fan-out plots (the x1-axis illustrates the absolute approach and the x2-axis illustrates
the relative approach) for the WikiSigned, Twitterlists and Brightkite
topologies. First row illustrates performance improvements regarding total time. Second row illustrates the effects of
this parameter on the network load.}
\label{fig:neighborandstopandfanout}
\end{figure*}
In terms of informed node percentages, only directed topologies have substantial benefits from larger communication fan-out
values. For instance, in the Twitterlists topology, setting $f_{abs}=3$, or $f_{rel}=4\%$, increases the average percentage
of informed nodes to 99.99\%, or 99.98\% respectively, up from 82.23\%. In general, the bare minimum fan-out
values needed to achieve high informed node percentages are $f_{abs}=3$, $f_{rel}=1\%$ and $f_{rel}=4\%$ since they deliver an
average informed node percentage of 99.35\%, 98.04\% and 100\% respectively. 

In Figure \ref{fig:neighborandstopandfanout}, we illustrate the effects of both communication fan-out approaches on the total
time and on the network load that is required to inform the nodes of three different types of topologies. The relative approach
performs better in signed and directed topologies, in terms of both total time and network load. For instance, in the WikiSigned
topology, by setting $f_{rel}=4\%$, we can reduce the rumour's spreading time, compared to the case of $f_{abs}=3$,
by 67.91\%, whilst also reducing the network load by 9.44\%. The absolute approach performs better in bidirectional topologies,
however, only in regards to total time. In the Brightkite topology, setting $f_{abs}=3$
reduces the rumour's spreading time by 33.34\%, compared to the case of $f_{rel}=4\%$, however, at the cost of increasing the
network load by 53.45\%. Across all topologies, $f_{rel}=4\%$ reduces the spreading time by an average of 42.98\%, whilst
$f_{abs}=3$ reduces it only by 18.60\%, both compared to the case of $f_{abs}=1$. Further increases in the fan-out parameter
improve the rumour's dissemination time, for each increment, by an average of 5.3\% for the absolute case (caps when $f_{abs}=13$),
and by an average of 10\% for the relative case (caps when $f_{rel}=13\%$). However, these small improvements do not justify the
additional load that is imposed on the network; for each increment of the fan-out parameter, the network load
is increased by 15.22\%, for
the absolute case, and by 20.92\% for the relative case, on average. Finally, we note that setting $f_{rel}=1\%$ does not
affect the rumour's spreading time, nor the load that is imposed on the network; it is only useful for boosting informed
node percentages.

The main concept of the relative approach is that it uses the many and useful links of popular nodes, which boost the rumour's
dissemination significantly, whilst keeping the traffic generated by unpopular nodes, whose links are of less value, to a minimum.
On the contrary, the absolute approach treats all links equally and attempts to balance the load across all nodes. This means
that the relative approach manages to spread the rumour throughout the core of the network much faster than the absolute approach,
regardless of the topology type.

The determinant factor, however, is still the convoy phenomenon, i.e., how chains of small communities, residing in the middle
region (group 2), can throttle the
protocol's performance, as illustrated in the beginning of the evaluation section. The benefit of increasing the communication
fan-out boosts the rumour's propagation throughout the chain only in bidirectional topologies, and only for
the absolute case, for two interrelated reasons. First, recall that the protocol can harness the power of both push and pull in
these topologies. Second, a simple increase of the absolute fan-out value can speed up dramatically the dissemination of
the rumour throughout the chain since hub nodes can now push the rumour to a neighboring community and, simultaneously, inform one of the nodes
in their community. Figure \ref{fig:neighborandstopandfanout3} illustrates clearly how drastically we can reduce the rumour's spreading
time simply by increasing the fan-out value to $f_{abs}=2$ and how further increases offer little improvement. 

The relative approach, however, does not scale as well as the absolute approach in bidirectional topologies. Middle region nodes, in their vast
majority, have really small out-degrees, which means that the value of $f_{rel}$ needs to be quite large ($\gg25\%$) in order for them
to communicate with more than one neighbor per round. Consequently, in bidirectional graphs, the relative approach is not capable
of improving the rumour's propagation throughout the chain as well as the absolute approach.

To summarize, in signed and directed topologies, a relative fan-out delivers the best improvements both in terms of total time
and network load. In bidirectional topologies, the best approach would be to combine the benefits of both the relative and absolute approaches.
Giant component nodes (group 3) should use a relative fan-out, while middle region nodes (group 2) should use an absolute fan-out.
\subsection{Enhanced Push \& Pull protocol}
\label{subsect:enhancedprotocol}
We conclude this section by presenting the performance improvements that an \emph{Enhanced push
\& pull} protocol
can provide compared to the vanilla asynchronous \emph{push \& pull} protocol. It combines all
of the knowledge that we have acquired from our experiments and it has the following parameter
values:

\begin{itemize}
	\item $O(\log^2{n})$ as its stopping policy.
	\item A relative communication fan-out $f_{rel}=4\%$. The only exception is in bidirectional
	topologies where middle region nodes have an absolute communication
	fan-out $f_{abs}=2$.
	\item Our novel QPU neighbor selection policy.
\end{itemize}

\begin{table}[ht!]
  \centering
  {
	\begin{tabular}{l c c c}
	\\ \hline
	\textbf{Network} & \textbf{90\% Inf.} & \textbf{97\% Inf.} & \textbf{100\% Inf.}
	\\ \hline
	Slashdot1 & 75.61\% & 87.82\% & 97.85\% \\
	Slashdot2 & 74.53\% & 87.45\% & 97.92\% \\
	Slashdot3 & 74.33\% & 87.37\% & 97.82\% \\
	Epinions & 73.31\% & 91.93\% & 99.59\% \\
	WikiSigned & 93.87\% & 97.40\% & 99.64\% \\
	Hamsterster & 50.43\% & 48.46\% & 59.12\% \\
	Brightkite & 52.26\% & 51.36\% & 59.94\% \\
	Facebook & 55.50\% & 53.05\% & 54.91\% \\
	TwitterLists & 79.73\% & 84.42\% & 93.95\% \\
	Google+ & 98.00\% & 98.75\% & 99.69\% \\
	\hline \hline
	Average & 72.76\% & 78.80\% & 86.04\%
	\end{tabular}
  }
	\caption{Percentage improvement of the total time to inform
	various node percentages that our \emph{Enhanced push \& pull} protocol achieves compared
	to the vanilla \emph{push \& pull} protocol.}
	\label{table:enhancedvsvanillastats}
\end{table}
Table \ref{table:enhancedvsvanillastats} illustrates the percentage improvement of the
total time to inform various node percentages that the \emph{enhanced} version of the 
protocol achieves compared to the standard vanilla version. The simple tweak of setting
the fan-out of the middle region nodes to $f_{abs}=2$, instead of $f_{rel}=4\%$, allows
us to spread the rumour fast throughout the core of the network and, at the same time,
successfully deal with the convoy phenomenon in bidirectional topologies, as was
previously illustrated. This tweak reduces the rumour's spreading time by an average of 18.74\%. However, we note that we experimented with higher values
as well. For instance, we find that by increasing the fan-out of the middle region nodes to $f_{abs}=3$, the rumour's
spreading time is reduced by an average of 6.96\%, compared to the case of $f_{abs}=2$,
however, at the cost of increasing the load on the network by an average of 24.03\%.
These minor improvements do not justify the additional load that is imposed on the network
and thus, $f_{abs}=2$ seems as the most suitable choice for the fan-out of the middle region nodes
in bidirectional topologies.

The biggest percentage improvements take place in directed and signed graphs. The best
result is in the Google+ topology where the total time required to inform all of the nodes
is reduced from an average of 20,238.75 seconds to an average of 62.73 seconds, which is
a percentage improvement of 99.69\%. Even in the average case, the data in the table suggest
how an intelligent selection of protocol parameters can deliver tremendous performance enhancements.

%% file: conclusionandfuture.tex
\section{Conclusions and Future Work}
\label{sect:conclusionsfw}

In this paper, we presented an in-depth experimental analysis of the
\emph{asynchronous push \& pull} rumour spreading protocol. We studied the behaviour of the protocol
over a large variety of real social network datasets (both common and signed).
This is the first study that examines how \emph{multiple} protocol parameters affect the
protocol's behaviour both in {\emph{isolation and combination} and under a wide range of values.

We illustrate how the convoy phenomenon, i.e., the inherent difficulty of the vanilla
protocol to randomly select isolated nodes that lie at the edge of the network,
can lead to extreme delays. However, when the goal is to inform large node percentages
(e.g. 97\%), the protocol manages to disseminate the rumour quickly. This can prove to
be extremely helpful for applications such as voting and quorum-based systems.


Our measurements indicate that a popular rumour originator can provide an improvement as high as
28.46\% in the total time required to inform 90\% of the nodes. Futhermore, we illustrate that
alternative neighbor contact criteria can provide significant improvements compared to the
standard approach of choosing uniformly at random. The best alternative is our novel QPU neighbor
selection policy, which is to cycle back and forth from popular to unpopular neighbors, yielding an average improvement
of 67.17\%.


By leveraging the knowledge attained from our empirical study, we also proposed an
\emph{Enhanced push \& pull} protocol that combines our QPU neighbor
selection policy, a relative communication fan-out $f_{rel}=4\%$ (in most cases) and $O(\log^2{n})$ as
its stopping policy. Our \emph{Enhanced push \& pull} protocol delivers an 86.04\% average percentage
improvement in the rumour's dissemination time over the plain-vanilla version. We believe
that our protocol provides strong evidence that rumours can indeed spread fast in real-world
social topologies when propagated intelligently.

Future work includes exploring protocol behaviour in cases where there are multiple
and distinct rumour originators that generate content dynamically. Note that this
problem is substantially different from the well-known and well-studied gossiping problem.